# Studies on an S-band bunching system with hybrid buncher

PEI Shi-Lun(裴士伦)[1)]   XIAO Ou-Zheng(肖欧正)

Institute of High Energy Physics, Chinese Academy of Sciences, Beijing 100049, China

**Abstract** Generally, a standard bunching system is composed by a SW pre-buncher, a TW buncher and a standard accelerating section. However, there is one way to simplify the whole system to some extent by using the hybrid buncher, which is a combined structure of the SW pre-buncher and the TW buncher. Here the beam dynamics studies on an S-band bunching system with the hybrid buncher is presented, simulation results shows that similar beam performance can be obtained at the linac exit by using this kind of bunching system rather than the standard one. In the meantime, the structure design of the hybrid buncher is also described. Furthermore, the standard accelerating section can also be integrated with the hybrid buncher, which can further simplify the usual bunching system and lower the construction cost.

## 1. Introduction

A standard bunching system is usually composed by a standing wave (SW) pre-buncher (PB), a travelling wave (TW) buncher and a standard accelerating section. However, for various reasons and different applications, the bunching system is often simplified to lower the construction cost or complicated to enhance the beam performance. In the industrial applications, the PB and buncher are often eliminated but the first few cells of the standard accelerating section are modified with gradually increasing phase velocity $\beta$; in the scientific applications, one or more Sub-harmonic Bunchers (SHB) are usually introduced into the bunching system or used to replace the PB [1]. The former is commonly accompanied with degraded beam performance; while the construction cost of the latter is relatively higher.

IHEP is constructing a 100 MeV / 100 kW electron linac (625 Hz / 2.7 μs / 600 mA) with the standard bunching system for the Kharkov Institute of Physics and Technology of National Science Center (NSC KIPT, Kharkov, Ukraine) [2]. Initially, the buncher is a 4-cell TW constant impedance (CI) structure with $\beta$=0.75. However, the adoption of water cooling jacket to bring away the high average RF power dissipated on the structure wall demands more longitudinal space (leading a longer buncher) to ease the installation. Finally, a 6-cell version was developed.

Inspired by the innovative idea of the hybrid photo-injector developed by the INFN-LNF/UCLA/SAPIENZA collaboration [3], we propose an alternative design to simplify the nominal/standard bunching system—replace the PB and B with the hybrid buncher, which is a combined structure of the PB and buncher. By using the hybrid bunching system, beam dynamics studies show that similar beam performance can be obtained at the KIPT linac exit but with a little bit lower construction cost.

## 2. Bunching system

Figure 1 shows the nominal/standard bunching system applied in KIPT linac and the alternative design we proposed to replace the PB and buncher by the hybrid buncher.

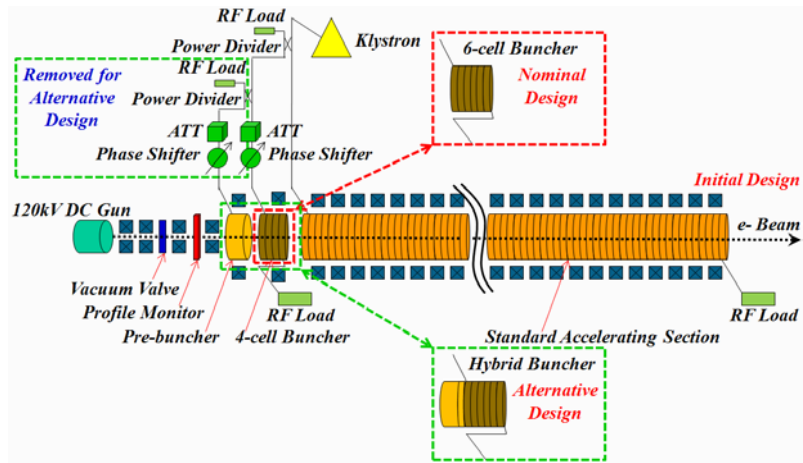

Figure 1: Bunching system.

1) E-mail: peisl@mail.ihep.ac.cn

It can be clearly seen from Fig. 1 that the hybrid bunching system has several advantages.

—More compact than a split system, allows scaling to higher frequency for a table-top system in the industrial applications.

—Much simpler high power RF system. Some waveguide section, attenuator, phase shifter, RF load can be avoided to lower the total construction cost.

—More flexible in the whole system installation and tuning, less parameters need to be adjusted and optimized.

—Completely removal of the impedance matching problem during the PB design and fabrication process, therefore the RF reflected power in the real operation.

However, the hybrid bunching system also has disadvantages.

—Slight degradation of the beam performance.

—Demands relatively accurate calculation of the longitudinal distance between the SW and TW sections, which is based on the gun emitted beam energy.

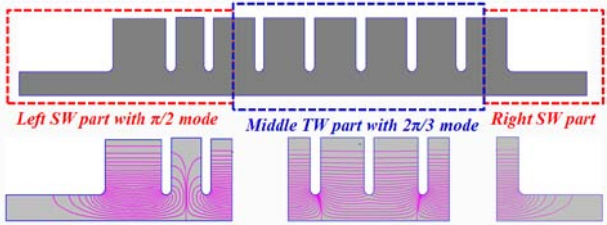

Figure 2: Initial 2D design of the hybrid buncher by SUPERFISH.

## 3. Initial 2D design of the hybrid buncher

To be comparable with the standard/split bunching system and check the feasibility of the hybrid bunching system in the KIPT linac, the TW section of the hybrid buncher was also designed to have 6 cells. To accommodate RF field data to PARMELA [4], the initial 2D design (shown in Fig. 2) of the SW and TW sections were performed separately by setting appropriate boundary conditions in SUPERFISH [5]. The SW section operates at $\pi/2$ mode, while TW section at $2\pi/3$. Fig. 3 shows the electric field amplitude distribution introduced into PARMELA for beam dynamics studies. After several iterations between the RF design and the dynamics simulation, the 2D design of the hybrid buncher can be finalized.

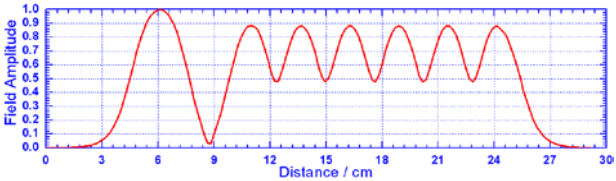

Figure 3: Electric field amplitude distribution for dynamics studies.

## 4. Beam dynamics simulation

To satisfy the energy spread requirement (<±4% p-to-p) at the 100 MeV / 100 kW linac exit [6], the hybrid bunching system should be able to produce similar energy spectrum as the nominal design, which is appropriate for the downstream collimation process realized by a dedicated Chicane system with a Collimator deployed. 600 mA electron beam (~70% efficiency) should be able to be obtained after collimation (at the chicane system exit). Based on the above considerations, the hybrid bunching system is optimized with the RF field data obtained from the 2D RF design. Fig. 4 shows the beam phase and energy spectrums at the hybrid bunching system exit, the corresponding spectrums for the nominal case are also shown for comparison. The hybrid bunching system has relatively lower transportation efficiency, while the bunch length of the main bunch part is relatively shorter.

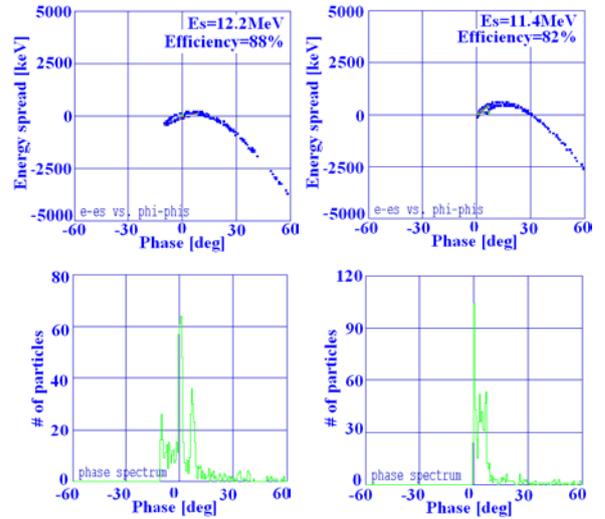

Figure 4: Spectrums at the bunching system exit (left for the nominal bunching system; right for the hybrid bunching system).

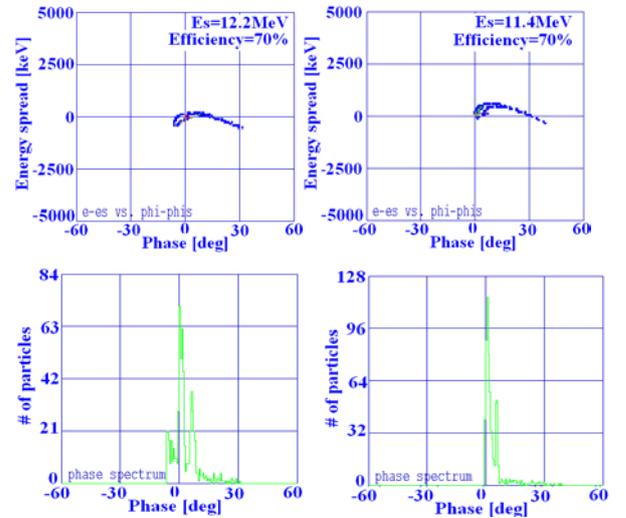

Figure 5: Spectrums at the chicane system exit (left for the nominal bunching system; right for the hybrid bunching system).

Figures 5 and 6 show the beam spectrums at the exits of the chicane system and the linac. Both the nominal and the hybrid bunching system can satisfy the transportation efficiency requirement from the electron gun to the linac exit. However, the hybrid bunching system produces relatively bigger absolute energy spread at the linac exit.

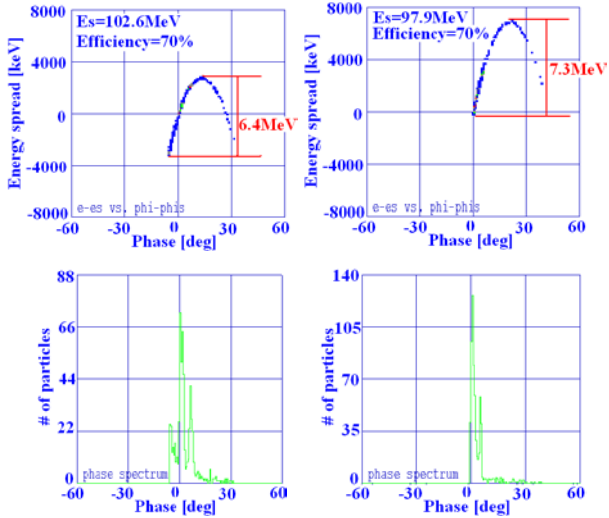

Figure 6: Spectrums at the linac exit (left for the nominal bunching system; right for the hybrid bunching system).

Fig. 7 shows the transverse emittance evolution along the linac. For the hybrid bunching system, the emittance at the linac exit is ~37% bigger, which is caused by the bunching process. The hybrid bunching system has a relatively bigger energy modulation by the SW section and a shorter drift space between the SW and TW sections. Certainly, the emittance can be reduced a little bit by optimizing the drift space length. However, it is unlikely to obtain the same emittance as the nominal bunching system; the drift space length can't be as long as the nominal case for RF power coupling reason, which means that a bigger energy modulation by the SW section is always needed.

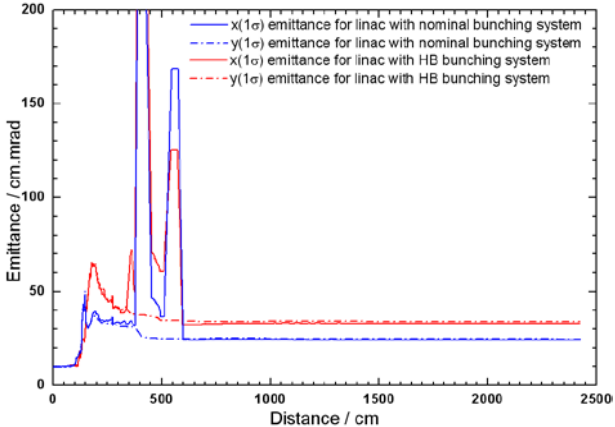

Figure 7: Transverse emittance evolution along the linac.

## 5. 3D design of the hybrid buncher

The 2D RF design of the Hybrid Buncher is the starting point of the 3D design. Similarly, the SW and TW sections of the hybrid buncher are designed separately. Fig. 8 shows the 3D model of the hybrid buncher, Table 1 lists the main dimensions.

Conceptually, the SW section can be further divided into two parts: the pre-bunching cavity and the coupling cavity. The former is used for beam energy/velocity modulation, while the latter as a beam drift. Both operate at $\pi/2$ mode. Length of each cavity is based on the 2D design and decided by the beam dynamics requirement. Based on the electron gun high voltage, the length of the coupling cavity in the SW section can be roughly calculated initially as the RF design baseline. Appropriate coupling cavity length is needed to place the beam at the positive slope of the RF field in the TW section to produce velocity bunching. The coupler matching of the TW section is based on the matching procedure for the $2\pi/3$ structure proposed by Dr. R. L. Kyhl and confirmed with the field transmission method. Attachment of the SW section with the TW section will increase the resonating frequency of the RF input coupler cavity a little bit; therefore a modest bigger cavity size is needed. In the meantime, the mesh size consistency in all of the 3D RF design steps should be kept to obtain correct result.

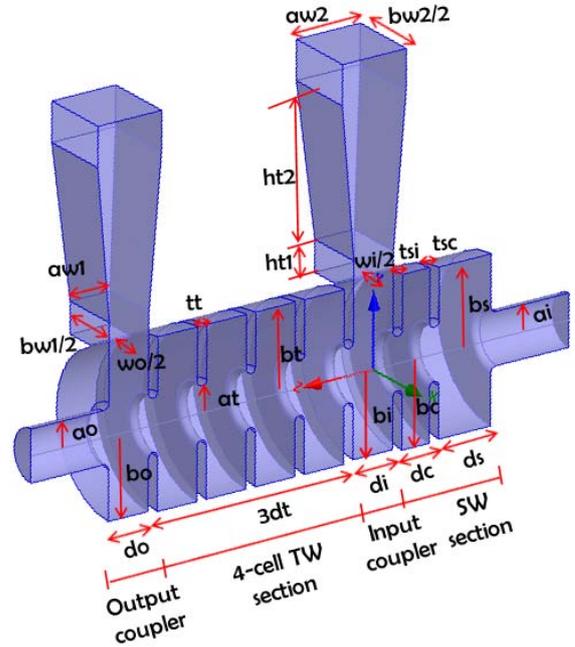

Figure 8: 3D model of the hybrid buncher.

Table 1: Main dimensions of the hybrid buncher

| Parameter | Value [mm] | Parameter | Value [mm] |
|---|---|---|---|
| ds | 32.803 | tsc, tsi, tt | 5 |
| dc | 19.682 | wi, wo | 30.08 |
| di, dt, do | 26.242 | ht1 | 15.9 |
| ai, at, ao | 13.11 | ht2 | 76.2 |
| bs | 41.556 | aw1 | 21.242 |
| bc | 42.442 | bw1 | 61.087 |
| bi, bo | 41.045 | aw2 | 34.163 |
| bt | 42.195 | bw2 | 72.263 |

Figure 9 shows the S11 curve at the RF input coupler port. Fig. 10 shows the phase and amplitude distribution of the RF electric field along the longitudinal axis of the hybrid buncher. To excite the RF field required by the beam dynamics, the total needed RF power is ~0.7 MW. Only ~3% of the RF power goes to the SW section, thus both the RF phase and the field amplitude in the TW section are not very sensitive to the resonate frequency of the SW section, which is different from the hybrid photo-injector [3]. <0.5° phase change is found when the SW section has an off-resonance of 100 kHz. For temperature control of ±0.1 °C (±5 kHz off-resonance), the corresponding phase variation caused by the frequency off-resonance of the SW section is ~±0.025°, which can be neglected in the real operation.

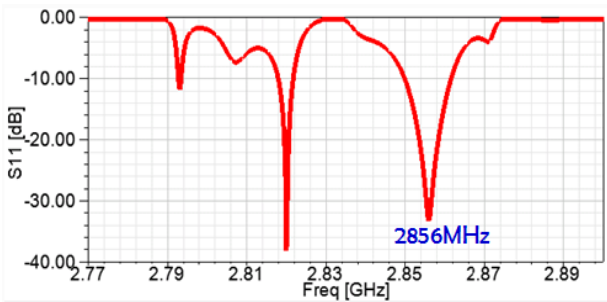

Figure 9: S11 curve at the RF input coupler port.

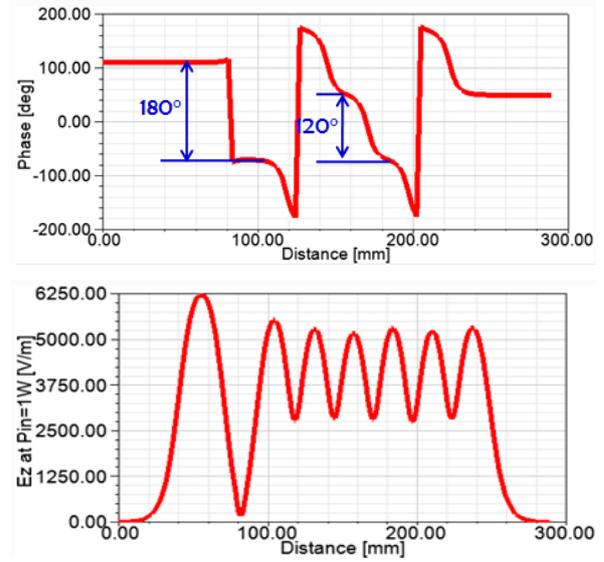

Figure 10: Phase (upper) and amplitude (lower) distribution of the RF electric field along the longitudinal axis of the hybrid buncher.

Figure 11 shows the typical RF electric field distribution applied in PARMELA and that produced by 3D structure design at different time shot. It can be seen that the electric field used in the dynamics simulation are consistent with that obtained with 3D RF design.

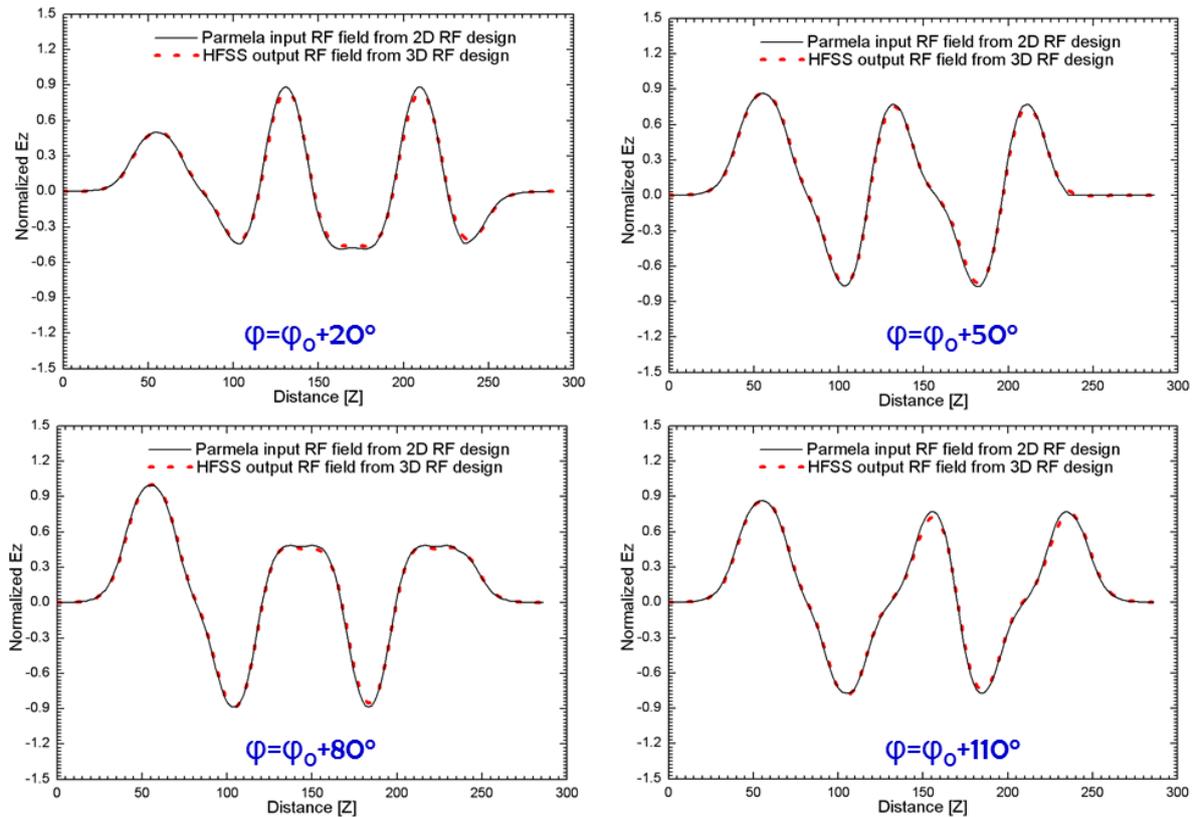

Figure 11: Typical RF electric field distribution at different time shot.

## 6. Summary

To simplify the standard bunching system with slightly degraded beam performance and relatively lower construction cost, a bunching system with hybrid buncher used to replace the PB and buncher was designed, which can still satisfy the linac design requirement.

For the electron gun high voltage quite different from 120 kV, the hybrid buncher needs to be redesigned.

Further optimization of the drift space length between the SW and TW sections can slightly reduce the transverse emittance.

Besides the integration of the PB and B, the standard accelerating section can also be integrated with the hybrid buncher together, which can further simplify the system and lower the construction cost.